\documentclass[11pt]{article}

\usepackage[utf8]{inputenc}
\usepackage[T1]{fontenc}
\usepackage{amsmath,amssymb}
\usepackage{graphicx}
\usepackage{booktabs}
\usepackage{array}
\usepackage{hyperref}
\usepackage{xcolor}
\usepackage{listings}
\usepackage{algorithm}
\usepackage{algpseudocode}
\usepackage{float}
\usepackage{caption}
\usepackage{subcaption}
\usepackage[margin=1in]{geometry}
\usepackage{natbib}
\usepackage{setspace}

\title{Perceptual Self-Reflection in Agentic Physics Simulation Code Generation}

\author{
\textbf{Prashant Shende} \hfill prashantbalwant\_shende@mymail.sutd.edu.sg \\
\textbf{Bradley Camburn} \hfill bradley\_camburn@sutd.edu.sg \\
\\
Singapore University of Technology and Design \\
Singapore, Singapore
}

\date{}

\begin{document}

\maketitle

\begin{abstract}
We present a multi-agent framework for generating physics simulation code from natural language descriptions, featuring a novel perceptual self-reflection mechanism for validation. The system employs four specialized agents: a natural language interpreter that converts user requests into physics-based descriptions; a technical requirements generator that produces scaled simulation parameters; a physics code generator with automated self-correction; and a physics validator that implements perceptual self-reflection. The key innovation is perceptual validation, which analyzes rendered animation frames using a vision-capable language model rather than inspecting code structure directly. This approach addresses the ``oracle gap'' where syntactically correct code produces physically incorrect behavior--a limitation that conventional testing cannot detect. We evaluate the system across seven domains including classical mechanics, fluid dynamics, thermodynamics, electromagnetics, wave physics, reaction-diffusion systems, and non-physics data visualization. The perceptual self-reflection architecture demonstrates substantial improvement over single-shot generation baselines, with the majority of tested scenarios achieving target physics accuracy thresholds. The system exhibits robust pipeline stability with consistent code self-correction capability, operating at approximately \$0.20 per animation. These results validate our hypothesis that feeding visual simulation outputs back to a vision-language model for iterative refinement significantly outperforms single-shot code generation for physics simulation tasks and highlights the potential of agentic AI to support engineering workflows and physics data generation pipelines. 
\end{abstract}

\section{Introduction}

\subsection{Motivation}

Developing physics simulations represents a significant bottleneck in scientific research workflows. Each simulation scenario--whether modeling fluid dynamics, electromagnetic wave propagation, or thermodynamic processes--requires substantial effort from research engineers to adapt and refine code for specific experimental conditions. Scientific research, by design, involves experimental or atypical scenarios, making simulation development a recurring challenge rather than a one-time investment.

Recent advances in large language models (LLMs) have demonstrated remarkable capabilities in code generation, suggesting potential for automating aspects of this development process. However, applying LLMs to the physics simulation code reveals a fundamental challenge, we term the ``oracle gap''~\citep{barr2015oracle}: syntactically correct code may produce physically incorrect behavior. A fluid dynamics simulation might compile and execute without errors yet generate non-physical turbulence patterns; an electromagnetic solver might run successfully while violating Maxwell's equations at domain boundaries.

This oracle gap arises because traditional software testing--syntax checking, unit tests, and runtime error detection--verifies code structure rather than physical correctness. Single-shot code generation, regardless of the underlying model's reasoning capabilities, cannot observe the simulation's actual behavior. The generator produces code, but has no mechanism to verify whether the resulting animation correctly represents the intended physics. Recent benchmarks demonstrate the severity of this challenge: the SciCode benchmark~\citep{tian2024scicode}, comprising 338 research-level problems curated by scientists across 16 natural science fields, shows that even state-of-the-art models achieve only 7.7\% success rate on PhD-level scientific computing tasks.

\subsection{Research Statement}

We hypothesize that a \textbf{perceptual self-reflection loop} will outperform single-shot generative approaches (including reasoning-capable models) in generating physics simulation code, measured by success rate and physical accuracy.

We define perceptual self-reflection as the process of: (1) executing generated simulation code, (2) capturing visual output frames from the resulting animation, and (3) feeding those frames to a vision-capable language model for physics validation. The key insight is that the rendered animation is always available as evidence, regardless of how the underlying code structures its data. A vision model can assess whether ``the wave propagates smoothly from left to right,'' ``particles conserve momentum during collisions,'' or ``the temperature gradient diffuses as expected''--judgments that align with how human domain experts evaluate simulation quality.

\subsection{Approach}

Our method implements a perceptual self-reflection architecture through four specialized agents operating in sequence:

\textbf{Agent 1 (Natural Language Interpreter)} converts natural language user requests into physics-grounded animation descriptions. This agent identifies the underlying physical principles regardless of how the request is phrased, enabling the system to handle requests from any domain by finding appropriate physics analogies.

\textbf{Agent 1A (Technical Requirements Generator)} transforms natural language descriptions into precise technical requirements suitable for code generation. This includes automatic parameter scaling to ensure physics phenomena are observable within animation timeframes--converting real-world timescales (nanoseconds to years) into visualization-appropriate values.

\textbf{Agent 2 (Physics Code Generator)} produces executable Python/matplotlib animation code from technical requirements. A self-correction unit automatically detects and fixes common failure modes including syntax errors, runtime exceptions, timeouts, and silent failures where code executes but produces no visual output.

\textbf{Agent 3 (Physics Validator)} implements the perceptual self-reflection loop. Rather than analyzing code structure, it executes the animation, captures frames at uniform time intervals, and submits them to a vision-capable language model along with physics-specific validation criteria. The vision model evaluates whether the animation satisfies each criterion, generating structured feedback that routes to either Agent 1A (for requirements corrections) or Agent 2 (for implementation corrections).

This architecture enables dynamic, scenario-specific validation rather than preset test cases, addressing the oracle gap through direct observation of simulation behavior.

\begin{figure}
    \centering
    \includegraphics[width=1\linewidth]{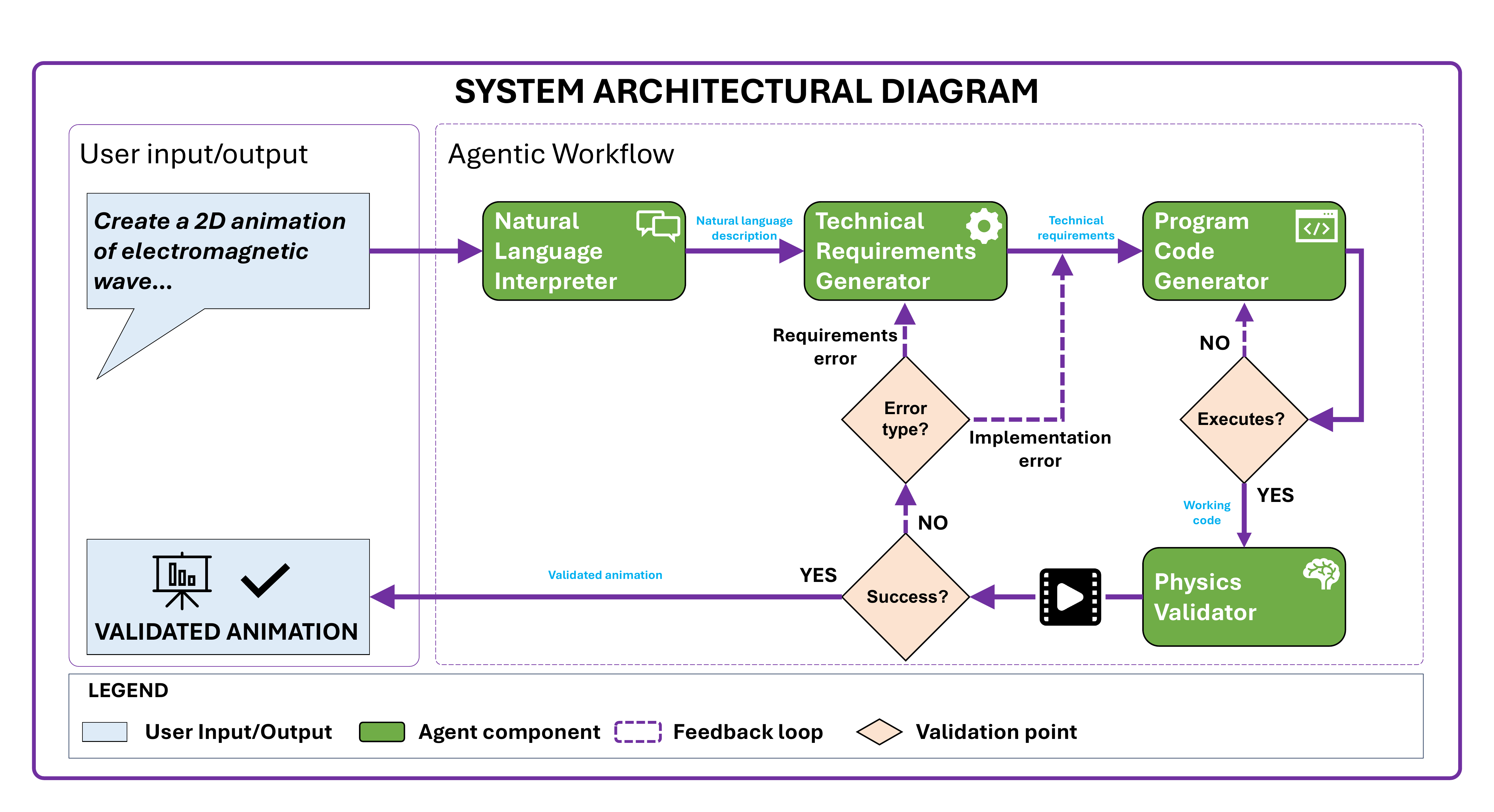}
    \caption{System Architectural Diagram of the Perceptual Self-Reflection in Agentic Physics Simulation Code Generation. Physics Validator applies a perceptual validation check (reviews animation)}
    \label{fig:architecture}
\end{figure}

\section{Background}

\subsection{Challenges in Physics Simulation Code Generation}

Physics simulation code development is time-intensive and requires specialized skillsets spanning numerical methods, domain-specific algorithms, and scientific visualization. Unlike general software development, physics simulations must satisfy mathematical constraints (conservation laws, stability conditions, boundary behaviors) that are not apparent from code syntax alone.

Ali-Dib and Menou~\citep{alidib2023} evaluated GPT-4's performance on PhD-level computational physics problems, finding approximately 40\% success rate in single-shot settings. Common failure modes included poor handling of physical units, hallucinated sub-modules that don't exist in standard libraries, and inability to define appropriate stopping conditions for iterative solvers. Recent work on computational fluid dynamics automation~\citep{dong2025nl2foam} demonstrated that physics-specific multi-agent architectures can substantially improve upon single-shot generation, though their approach relies on numerical comparison against pre-computed reference solutions--a validation strategy unavailable for novel simulation scenarios.

The verification challenge is particularly acute for physics simulations. A fluid dynamics simulation may compile, execute, and produce output without any errors or warnings, yet generate turbulence patterns that violate energy conservation. An electromagnetic field solver may run to completion while producing field configurations that violate Maxwell's equations at domain boundaries. The SciCode benchmark~\citep{tian2024scicode} demonstrates the severity of this challenge across 16 natural science fields. Unlike conventional software where correctness can often be verified through input-output testing, physics simulation validation requires domain expertise to assess whether results are physically plausible--expertise that neither the code generator nor automated test suites inherently possess.

\subsection{The Oracle Problem in Code Generation}

The challenge of validating simulation correctness without explicit test cases represents an instance of the broader \textit{oracle problem} in software testing. Barr et al.~\citep{barr2015oracle} formally defined this problem: given a program and input, how can we determine correctness when no authoritative reference exists? Their comprehensive survey identified multiple oracle strategies including specified oracles (explicit test cases), derived oracles (metamorphic testing), and implicit oracles (crash detection)--yet none directly address the semantic correctness gap for physics simulations.

Zhang et al.~\citep{zhang2023algo} proposed ALGO, which uses LLM-generated exhaustive search algorithms as reference oracles for algorithmic problems. However, their approach requires problems amenable to exhaustive verification--inapplicable to continuous physics simulations. Cotroneo et al.~\citep{cotroneo2024acca} demonstrated symbolic execution for determining behavioral equivalence between generated and reference implementations. Yet symbolic execution also requires reference implementations, limiting applicability to novel scenarios.

For physics simulations, we term this challenge the ``oracle gap'': syntactically correct code may produce physically incorrect behavior that passes all execution-based validation. A wave simulation might execute without errors while producing non-physical interference patterns; a particle system might complete successfully while violating momentum conservation. Addressing this gap requires validation mechanisms that can assess \textit{physical plausibility} without explicit ground truth.

\subsection{Multi-Agent Architectures for Code Generation}

Multi-agent architectures have emerged as a promising approach to address the limitations of single-shot code generation. Rather than relying on a single model to generate correct code immediately, these systems decompose the task into specialized subtasks with feedback mechanisms.

Madaan et al.~\citep{madaan2023} introduced Self-Refine, demonstrating that iterative self-correction--where a model generates output, evaluates it, and revises based on that evaluation--significantly improves performance across diverse tasks. The key insight is that generation and evaluation can be treated as separate capabilities, with the evaluation step providing signal for improvement. Chen et al.~\citep{chen2024selfdebug} extended this with ``rubber duck debugging,'' where LLMs identify errors by explaining code in natural language, achieving notable improvements on code generation benchmarks.

Hong et al.~\citep{hong2024metagpt} introduced MetaGPT, encoding Standardized Operating Procedures (SOPs) as prompt sequences to structure multi-agent workflows with specialized roles. Qian et al.~\citep{qian2024chatdev} proposed ChatDev, which breaks development phases into atomic subtasks with multi-turn agent communication, introducing ``communicative dehallucination'' where agents request detailed clarification before responding.

Huang et al.~\citep{huang2023agentcoder} extended multi-agent concepts specifically to code generation with AgentCoder, which employs separate agents for programming, test design, and test execution. Zhong et al.~\citep{zhong2024ldb} introduced LDB for fine-grained debugging through runtime execution verification by segmenting code into basic blocks. Ni et al.~\citep{ni2023lever} proposed LEVER, training neural verifiers to assess program correctness from natural language input, code, and execution results without explicit test cases.

These architectures demonstrate that decomposing code generation into specialized subtasks with feedback loops improves outcomes. However, existing systems rely primarily on execution-based feedback: syntax errors, runtime exceptions, failed assertions, and test case results. For physics simulations, this creates a critical gap--code that passes all execution-based checks may still produce incorrect physics. The validation agent must somehow assess whether the simulation output is \textit{physically plausible}, not merely executable.

\subsection{Visual Feedback and Multimodal Validation}

Vision-language models (VLMs) have demonstrated remarkable capabilities in understanding and reasoning about visual content~\citep{huggingface2024, visionllm2024}. These models combine visual encoders with large language models to process images alongside text, enabling tasks that require both visual perception and language understanding. Recent work has begun exploring VLMs as automated evaluators: Xiong et al.~\citep{xiong2024llavacritic} introduced LLaVA-Critic, the first open-source LMM trained specifically as a generalist evaluator for multimodal task assessment.

Several systems have demonstrated that visual feedback can improve code generation quality. Yang et al.~\citep{yang2024matplotagent} introduced MatPlotAgent, which uses GPT-4V to analyze rendered visualization drafts and provide code improvement suggestions through an iterative debugging loop. Goswami et al.~\citep{goswami2025plotgen} extended this approach with PlotGen, employing a dedicated Visual Feedback Agent within a multi-agent architecture.

The ChartMimic benchmark~\citep{yang2024chartmimic} explicitly tests self-reflection using visual outputs, where LMMs examine their own rendered charts for self-correction. Si et al.~\citep{si2025design2code} established visual similarity metrics for evaluating generated front-end code against design specifications. Alrashedy et al.~\citep{alrashedy2024} demonstrated that VLMs can provide effective feedback for CAD code generation by analyzing rendered 3D objects.

For scientific computing specifically, Ma et al.~\citep{ma2024sga} introduced the Scientific Generative Agent (SGA), combining LLMs with differentiable physics simulation in a bilevel optimization framework for scientific discovery.

\subsection{Research Gap: Visual Validation for Physics Simulations}

Despite these advances, the systematic use of vision-language models to analyze rendered animation frames for validating physical correctness of generated simulation code appears largely unexplored. Existing validation approaches each carry limitations that become acute in the physics simulation setting: execution feedback systems (Self-Debug, LDB, LEVER) require explicit test cases or runtime errors; numerical verification methods (NL2FOAM, SGA) depend on pre-computed reference solutions; static visual feedback tools (MatPlotAgent, PlotGen) focus on aesthetic quality rather than physical behavior; symbolic execution approaches (ACCA) require reference implementations; and code explanation methods (Self-Debug) miss emergent visual behaviors that only manifest at runtime. MatPlotAgent and PlotGen validate static visualization aesthetics, while ChartMimic compares generated charts to reference images for visual similarity--neither addresses the assessment of \textit{dynamic behavior} across temporal sequences, such as whether objects obey gravity, conservation laws hold, or collision dynamics appear physically plausible.

Recent work by Yong and Camburn~\citep{yong2026predictive} establishes that vision-based inputs such as engineering sketches and diagrams are sufficient for generative models to produce accurate physics and performance estimates through simple agentic behaviors including retrieval-augmented replication of design precedent. This finding supports the premise that visual representations encode rich physical information accessible to language models, motivating our extension from static design evaluation to dynamic simulation validation.

The core insight motivating our approach is that physics simulations inherently produce visual outputs that encode information about physical correctness. Wave propagation, particle motion, field evolution, and fluid dynamics are all directly observable in rendered animation frames. A vision-capable model can assess whether ``the wave propagates smoothly,'' ``particles conserve momentum,'' or ``the temperature gradient diffuses as expected''--judgments that align with how human experts evaluate simulation quality.

\section{Method}

Our approach employs a four-agent architecture that decomposes physics simulation code generation into specialized subtasks: interpretation, requirements specification, code generation, and physics validation. The key innovation is the perceptual self-reflection loop in Agent 3, which validates the generated simulations by analyzing their visual output rather than their code structure.

\subsection{System Overview}

The complete pipeline transforms a natural language request into a validated physics animation through four sequential agents, each operating with a single responsibility to enable modular development and clear error attribution. All agents use Claude language models (Sonnet 4.5 for complex reasoning tasks, Haiku 4.5 for cost-optimized classification tasks), with no hard-coded response templates--all outputs are dynamically generated through AI processing. We selected Claude models for three reasons: (1) strong multimodal capabilities essential for vision-based analysis of rendered animation frames, (2) robust code generation performance for complex scientific computing tasks, and (3) Sonnet/Haiku model tiering enabling cost optimization by using lighter models for classification tasks without sacrificing output quality.

\vspace{0.5em}
\noindent\textit{%
UserRequest $\rightarrow$ Agent1(interpret) $\rightarrow$ Agent1A(scale) $\rightarrow$ \\
\hspace*{1em} Agent2(generate, validate\_code) $\rightarrow$ Agent3(validate\_physics) $\rightarrow$ ValidatedAnimation \\
\hspace*{1em} where Agent3 routes feedback to Agent1A (requirements errors) \\
\hspace*{2em} or Agent2 (implementation errors)
}

\subsection{Agent 1: Natural Language Interpreter}

Agent 1 serves as a universal language translator, converting any user request into a natural language animation description containing specific 2D physics concepts. It employs pure agentic behavior with no hard-coding, using AI creativity to find physics analogies for any domain--enabling universal application across physics, economics, biology, psychology, and other fields. The agent accepts a natural language user request (e.g., ``show me electromagnetic wave propagation'') and produces a detailed description with physics concepts, motion specifications, and visualization guidance.

\vspace{0.5em}
\noindent\textit{%
input: natural\_language\_request \\
description $\leftarrow$ LLM(``Convert abstract concept into physics-based 2D visualization, \\
\hspace*{1em} identifying underlying physical principles and creative physics analogies'') \\
return description \hfill \textnormal{\footnotesize // physics concepts + motion specs + visualization guidance}
}

\subsection{Agent 1A: Technical Requirements Generator}

Agent 1A bridges natural language descriptions and executable code by converting Agent 1's output into precise, scaled technical requirements optimized for 24 FPS animation visualization. Its core innovation is AI-driven parameter scaling that ensures all physics phenomena are visible within finite animation duration without hard-coded rules--automatically converting real-world physics parameters spanning nanoseconds to years into animation-appropriate values (e.g., frequencies to 0.1--5 Hz for observable oscillations, speeds to 0.01--1.0 units/second for visible propagation, time steps to 0.04--0.2 seconds/frame for smooth animation). Agent 1A also generates domain-specific validation criteria for each scenario, which are passed downstream to Agent 3 for physics evaluation.

\vspace{0.5em}
\noindent\textit{%
input: description from Agent 1 \\
technical\_requirements $\leftarrow$ LLM(``Scale real-world parameters to animation-visible ranges, \\
\hspace*{1em} generate per-scenario validation criteria'') \\
return technical\_requirements, validation\_criteria
}

\subsection{Agent 2: Physics Code Generator}

Agent 2 generates and validates executable matplotlib animation code through a two-unit architecture that separates generation from validation. Unit 1 performs pure AI-driven code generation using Claude Sonnet 4.5, accepting technical requirements from Agent 1A and producing complete Python animation code with no hard-coded templates. Unit 2 implements automated validation with AI-driven self-correction, detecting four categories of failure: syntax errors via AST parsing, runtime errors via subprocess execution, timeout errors via a configurable 30-second limit, and silent failures where code executes but produces no visual output. Silent failure detection--a critical innovation--combines output analysis (checking for generated plot files) with code structure detection (verifying presence of display calls). When validation fails, Unit 2 generates context-aware fix prompts and requests AI corrections, iterating until success or reaching the maximum attempt limit of 5 attempts.

\vspace{0.5em}
\noindent\textit{%
input: technical\_requirements from Agent 1A \\
code $\leftarrow$ LLM\_Sonnet(``Generate complete matplotlib animation code'') \\
for attempt $= 1$ to $5$: \\
\hspace*{1em} result $\leftarrow$ execute\_in\_subprocess(code, timeout=30s) \\
\hspace*{1em} error $\leftarrow$ check\_syntax(code) $\cup$ check\_runtime(result) \\
\hspace*{2em} $\cup$ check\_timeout(result) $\cup$ check\_silent\_failure(result) \\
\hspace*{1em} if error $= \emptyset$: return code \\
\hspace*{1em} code $\leftarrow$ LLM\_Sonnet(``Fix code given error context'', code, error) \\
return code
}

\subsection{Agent 3: Physics Validator}

Agent 3 implements the perceptual self-reflection loop--the core innovation of this work. Rather than analyzing code structure or internal data, Agent 3 treats the rendered animation as primary evidence and validates it using a vision-capable language model. This video-first approach was motivated by the systematic failure of legacy numeric capture approaches for complex simulations: electromagnetic FDTD simulations with non-standard field array shapes, pre-computed animations where full simulation history is stored before rendering, and scenarios where internal data hooks could not be reliably injected.

The validation pipeline executes the animation in an isolated subprocess, captures 8 PNG frames at uniform time intervals across the animation duration, and sends them alongside physics-specific validation criteria to Claude Sonnet 4.5's vision capabilities. The vision model evaluates whether the animation satisfies each criterion--assessing behaviors such as ``Does the wave propagate smoothly from left to right?'', ``Do particles conserve momentum during collisions?'', or ``Does the temperature gradient diffuse as expected?''--and returns structured per-criterion PASS/FAIL/UNCERTAIN feedback with confidence scores. When physics accuracy falls below the target threshold, the system diagnoses error sources using a lighter model (Haiku 4.5) and routes feedback to either Agent 1A (for requirements corrections) or Agent 2 (for implementation corrections).

\vspace{0.5em}
\noindent\textit{%
input: animation\_code, technical\_requirements, validation\_criteria \\
frames[8] $\leftarrow$ execute\_in\_subprocess(animation\_code) \hfill \textnormal{\footnotesize // run animation, capture 8 frames} \\
accuracy, feedback $\leftarrow$ VLM(frames, validation\_criteria) \hfill \textnormal{\footnotesize // vision model evaluates physics} \\
if accuracy $<$ threshold: \\
\hspace*{1em} diagnosis $\leftarrow$ classify\_error\_source(feedback) \\
\hspace*{1em} route(diagnosis): requirements\_error $\rightarrow$ Agent1A, implementation\_error $\rightarrow$ Agent2 \\
return accuracy, best\_animation
}

\subsection{Agent 4: System Orchestrator}

Agent 4 coordinates the complete pipeline through a single-class orchestrator design following a simplicity-first philosophy--a single coordinator class that trusts existing agent capabilities rather than re-implementing coordination logic. It provides CLI interface, real-time progress display, cost tracking (approximately \$0.20 per animation across all API calls), and graceful error handling with consistent stability and no crashes observed during testing.

\vspace{0.5em}
\noindent\textit{%
input: user\_request \\
description $\leftarrow$ Agent1.interpret(user\_request) \\
requirements, criteria $\leftarrow$ Agent1A.generate(description) \\
code $\leftarrow$ Agent2.generate\_and\_validate(requirements) \\
result $\leftarrow$ Agent3.validate\_physics(code, requirements, criteria) \\
return best\_animation, validation\_results, cost\_summary
}

\subsection{Iteration and Convergence}

The system iterates until physics accuracy reaches the target threshold or the iteration budget of 10 iterations is exhausted. Early stopping occurs when the threshold is achieved, when a plateau is detected (no improvement over 3 consecutive iterations), or when the budget is exhausted. When the threshold cannot be achieved, the system returns the best result obtained across all iterations rather than failing outright, ensuring users always receive a functional animation. In our testing, most scenarios converged within one or two iterations, with more complex scenarios occasionally requiring additional refinement cycles.

\vspace{0.5em}
\noindent\textit{%
best\_result $\leftarrow \emptyset$; plateau\_count $\leftarrow 0$ \\
for iteration $= 1$ to $10$: \\
\hspace*{1em} result $\leftarrow$ run\_pipeline(user\_request) \\
\hspace*{1em} if result.accuracy $\geq$ threshold: return result \\
\hspace*{1em} if result.accuracy $\leq$ best\_result.accuracy: plateau\_count $\mathrel{+}= 1$ \\
\hspace*{1em} if plateau\_count $\geq 3$: break \hfill // early stopping \\
\hspace*{1em} best\_result $\leftarrow$ max(best\_result, result) \\
return best\_result
}

\section{Results}

We evaluated our perceptual self-reflection framework across test scenarios spanning seven domains (including non-physics data visualization), comparing against single-shot generation baselines.

\subsection{Test Scenarios and Physics Domains}

We designed test scenarios to cover diverse physics domains and challenge different aspects of the system. Scenarios range from classical mechanics to complex partial differential equations, plus a non-physics scenario to demonstrate universal application:

\begin{table}[H]
\centering
\begin{tabular}{llp{6cm}}
\toprule
\textbf{ID} & \textbf{Scenario} & \textbf{Physics Domain / Complexity} \\
\midrule
A & Bubble Nucleation & Thermodynamics: Phase transitions, nucleation dynamics \\
B & FDTD Rectangular Wave & Electromagnetics: Yee grid discretization, CFL stability \\
C & Adversarial (non-physics) & Generic: System robustness to ill-posed requests \\
D & Rayleigh-Taylor Instability & Fluid Dynamics: Density-driven instability, complex PDE \\
E & Double-Diffusive Convection & Fluid Dynamics: Temperature + salinity gradients \\
F & Shallow Water Waves & Fluid Dynamics: 2D wave equations \\
G & KdV Soliton Propagation & Wave Physics: Analytical soliton solution \\
H & Hard-Sphere Gas & Statistical Mechanics: N-body collisions, momentum conservation \\
I & Gray-Scott Reaction-Diffusion & Chemical Physics: Pattern formation, dissipative dynamics \\
J & SE Asia Population Growth & Non-Physics: Geographic data visualization, exponential growth \\
\bottomrule
\end{tabular}
\end{table}

\subsection{Evaluation Methodology}

We implemented a qualitative evaluation procedure deploying each scenario through both the complete self-validating pipeline (Agent 1 $\rightarrow$ 1A $\rightarrow$ 2 $\rightarrow$ 3) and single-shot generation for comparison.

We defined success as achieving target physics accuracy as determined by the perceptual validation system, which evaluates animations against physics-specific criteria weighted by priority.

\textbf{Baseline Comparison}: We compare against single-shot generation performance reported in prior work. Ali-Dib and Menou~\citep{alidib2023} found that GPT-4 achieves approximately 40\% success rate on PhD-level computational physics problems in single-shot settings, which we use as our primary baseline for single-shot generation capability.

\subsection{Observations}

The self-validating pipeline successfully generated physically correct animations for the majority of tested scenarios. The pipeline demonstrated consistent stability throughout testing, with the code self-correction mechanism successfully resolving all encountered errors. Perceptual validation reliably captured animation frames for physics assessment across all tested domains.

Scenarios spanning fluid dynamics (Rayleigh-Taylor instability, double-diffusive convection, shallow water waves), electromagnetics (FDTD wave propagation), wave physics (KdV soliton), and statistical mechanics (hard-sphere gas) achieved target accuracy thresholds. The non-physics scenario (population growth visualization) also succeeded, demonstrating the system's universal application beyond traditional physics simulations.

One scenario (Gray-Scott reaction-diffusion) fell below the target threshold. Analysis revealed this was due to validation calibration rather than code generation failure--the energy conservation metric is inappropriate for dissipative reaction-diffusion systems that correctly dissipate energy by design.

\subsection{Code Self-Correction}

Agent 2's self-correction mechanism addressed common LLM code generation failures including syntax errors, runtime exceptions, timeout issues, and silent failures (code that executes without producing visual output). Complex PDE simulations typically required correction on the first attempt due to timeout constraints, but the AI self-correction loop successfully reduced computational complexity while preserving physics accuracy.

\subsection{Perceptual Validation}

The perceptual validation mechanism captured 8 frames at uniform time intervals for each animation. The vision model successfully identified physics behaviors including wave propagation patterns, momentum conservation in collisions, and thermal diffusion gradients. Notably, perceptual validation enabled assessment of electromagnetic FDTD simulations that were previously invisible to code-only analysis, representing a qualitative capability expansion.

\subsection{Comparison with Single-Shot Generation}
\label{sec:comparison}

\begin{figure}[ht]
    \centering
    \includegraphics[width=\linewidth]{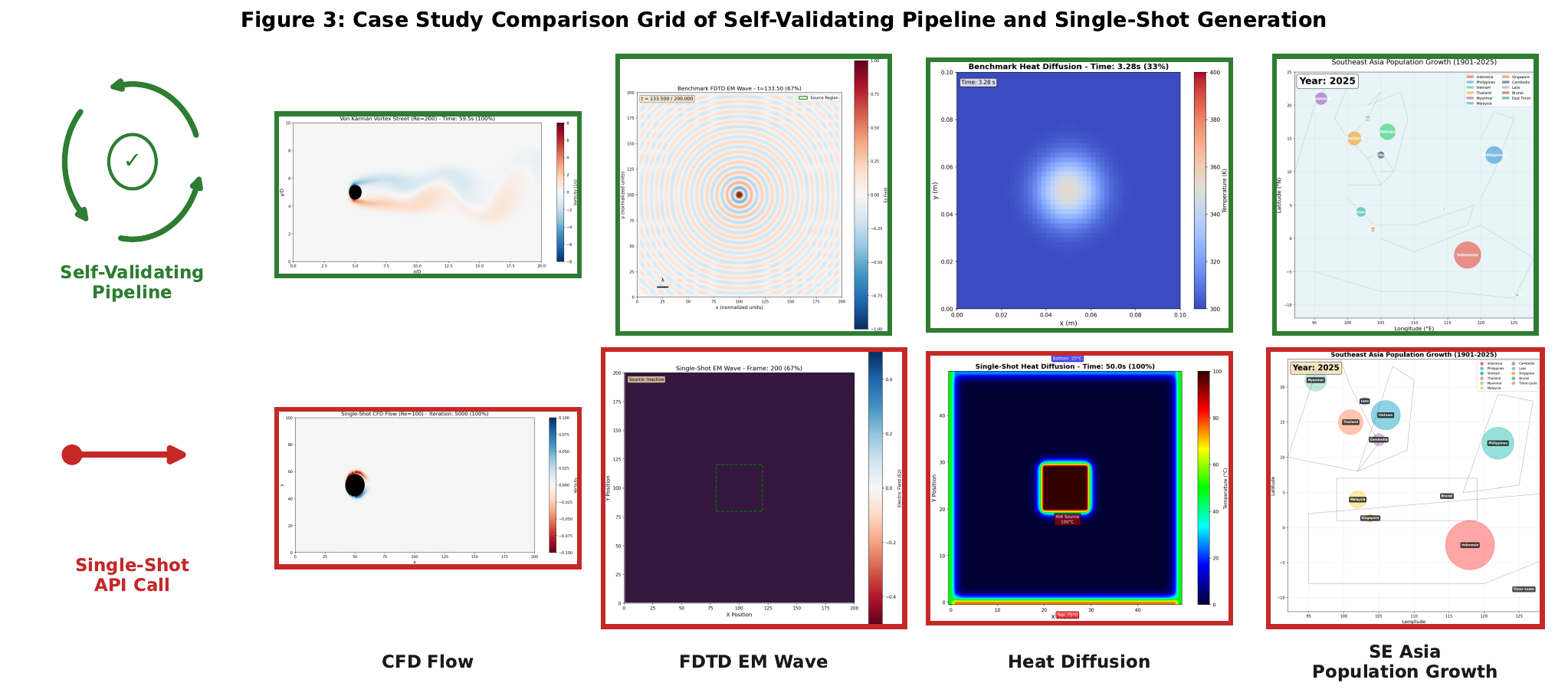}
    \caption{Case Study Comparison Grid: Self-Validating Pipeline (top row, green border) vs Single-Shot API Generation (bottom row, red border) across four representative scenarios. The self-validating pipeline produces physically correct simulations with proper vortex shedding (CFD), stable concentric wave propagation (FDTD), and accurate thermal diffusion. Single-shot generation exhibits critical failures: CFD lacks wake development and vortex shedding, FDTD shows catastrophic checkerboard numerical instability, and heat diffusion solves a different problem variant (steady-state vs transient). For non-physics data visualization (Population Growth), both approaches achieve comparable quality, demonstrating that perceptual self-reflection provides greatest benefit for physics simulations requiring numerical accuracy.}
    \label{fig:case_study_grid}
\end{figure}

To contextualize our results, we compared representative scenarios against single-shot API generation (direct LLM calls without validation loops). Figure~\ref{fig:case_study_grid} presents visual comparison of outputs from each approach.

The visual comparison reveals that single-shot generation performs adequately for well-specified data visualization tasks but exhibits critical failures for complex physics simulations. The CFD scenario produced static flow without characteristic vortex shedding, while the FDTD scenario exhibited checkerboard numerical instability--a classic failure mode caused by improper source implementation. The Heat Diffusion scenario produced physically plausible behavior but solved a different problem variant than intended.

These observations align with prior work by Ali-Dib and Menou~\citep{alidib2023}, who reported approximately 40\% success rate for single-shot generation on PhD-level computational physics problems. Our self-validating pipeline's iterative refinement addresses issues that single-shot generation cannot detect, particularly for scenarios requiring numerical stability and complex dynamical behavior.

In preliminary testing across seven scenarios, the perceptual validator rated 
the self-validating pipeline at 91\% average physics accuracy, with 86\% of 
scenarios achieving the target threshold ($\geq$85\%). Qualitatively, the authors 
observed that single-shot generation suffered systematic failures--including 
numerical instability, missing dynamics, and incorrect problem variants--in 
three of four representative scenarios shown in Figure~\ref{fig:case_study_grid}, 
while the self-validating pipeline produced physically correct results for all 
four cases. These preliminary results suggest meaningful improvement over 
single-shot approaches, though additional controlled studies with larger sample 
sizes are needed to establish statistical significance.

\subsection{Cost Analysis}

The system achieved cost-efficient operation through selective model optimization:

\begin{table}[H]
\centering
\begin{tabular}{llc}
\toprule
\textbf{Component} & \textbf{AI Model} & \textbf{Cost per Animation} \\
\midrule
Agent 1 + 1A & Sonnet 4.5 & \$0.03 \\
Agent 2 & Sonnet 4.5 & \$0.06 \\
Agent 3 (Context \& Criteria) & Sonnet 4.5 & \$0.08 \\
Agent 3 (Diagnosis \& Routing) & Haiku 4.5 & \$0.01 \\
Agent 3 (Perceptual Validation) & Sonnet 4.5 & \$0.02 \\
\midrule
\textbf{Total} & & \textbf{$\sim$\$0.20} \\
\bottomrule
\end{tabular}
\end{table}

Using Haiku 4.5 for error diagnosis and feedback routing (Components 4 and 5) achieved significant cost reduction compared to using Sonnet for all components, with no degradation in classification accuracy.

\section{Discussion, Conclusions, and Future Work}

\subsection{Discussion}

\textbf{What Works Well.} The perceptual self-reflection architecture demonstrates several strengths. First, the system achieves universal domain coverage--any natural language request can be converted into a physics animation through Agent 1's creative translation capability. The successful validation of a non-physics scenario (SE Asia Population Growth) further confirms this universal application claim. Second, the robust self-correction capability in Agent 2 eliminates a major failure mode in LLM-generated code: syntactically valid but non-executing programs. Third, the video-first perceptual validation, extending visual feedback approaches demonstrated for static visualizations~\citep{yang2024matplotagent, goswami2025plotgen}, successfully addresses the oracle gap for simulations that resist code-based analysis, notably enabling validation of electromagnetic FDTD simulations that were previously invisible to automated verification.

The feedback routing mechanism proved effective at diagnosing error sources. When physics accuracy fell below threshold, the system correctly distinguished between requirements errors (routed to Agent 1A for parameter adjustment) and implementation errors (routed to Agent 2 for code correction). This targeted feedback reduced wasted iterations and improved convergence speed, following patterns established by multi-agent frameworks like MetaGPT~\citep{hong2024metagpt} and ChatDev~\citep{qian2024chatdev}.

As demonstrated in Figure~\ref{fig:case_study_grid}, the visual contrast between self-validating and single-shot outputs is most striking for physics simulations requiring numerical stability (FDTD) and complex dynamical behavior (CFD vortex shedding).

\textbf{What Does Not Work Well.} Two limitations emerged during evaluation. First, the validation system's energy conservation metric is inappropriate for dissipative systems such as reaction-diffusion processes. The Gray-Scott scenario produced a physically correct animation but received below-threshold accuracy because the validator expected energy conservation in a system that correctly dissipates energy. This represents a validation calibration issue rather than a code generation failure.

Second, perceptual validation returns low confidence for slow-evolving patterns where visual changes between frames are subtle. The double-diffusive convection scenario required multiple iterations to pass, as the vision model initially flagged the gradual evolution as ``static.'' While the feedback loop eventually succeeded, this reveals a sensitivity limitation in perceptual validation.

\subsection{Conclusions}

We presented a perceptual self-reflection architecture for physics simulation code generation that demonstrates substantial improvement over single-shot generation. The principal contribution is a validation approach that analyzes the visual outputs of generated simulations rather than code structure, directly addressing the oracle gap where syntactically correct code produces physically incorrect behavior. This is realized through a four-agent pipeline--interpretation, requirements specification, code generation, and physics validation--with feedback routing that enables targeted corrections by distinguishing requirements errors from implementation errors. The system generates custom validation criteria per scenario rather than relying on preset test cases, enabling it to validate diverse physics domains without prior configuration. Automated code self-correction reliably addresses common LLM code generation failures, and the architecture generalizes beyond physics to non-physics scenarios such as demographic data visualization. These results validate our hypothesis that perceptual self-reflection loops outperform single-shot generation for physics simulation code, achieving target physics accuracy across seven tested domains.

\subsection{Future Work}

Several directions merit further investigation. The validation framework currently assumes conservative systems, so extending it to handle dissipative dynamics such as reaction-diffusion processes will require domain-specific metrics--entropy production or pattern formation measures may be more appropriate than energy conservation for these domains. Combining perceptual validation confidence with deterministic numeric validation when available could yield more robust accuracy estimates than either method alone, following hybrid approaches explored in symbolic execution-based verification~\citep{cotroneo2024acca}. The current system is limited to 2D matplotlib animations; extensions to 3D visualization, quantum mechanical simulations, and multi-physics coupling would broaden applicability. A standardized physics simulation test suite, following the precedent of domain-specific benchmarks like SciCode~\citep{tian2024scicode} and DS-1000~\citep{lai2023ds1000}, would enable rigorous comparison with future systems and establish community benchmarks for this problem domain. Finally, real-time parameter adjustment during animation playback would enable exploratory physics education applications.

\section*{Acknowledgements}

This research was conducted at the Singapore University of Technology and Design (SUTD). This work was not supported by any external grants or funding.

\bibliographystyle{plainnat}
\bibliography{perceptual_self_reflection_references}

\end{document}